# Signature of superradiance from a nitrogen gas plasma channel produced by strong field ionization

Guihua Li,[1,3,4] Chenrui Jing,[1,3,4] Bin Zeng,[1] Hongqiang Xie,[1,3] Jinping Yao,[1] Wei Chu,[1] Jielei Ni,[1] Haisu Zhang,[1,3] Huailiang Xu,[2,†] Ya Cheng[1,*] and Zhizhan Xu [1,#]

[1] *State Key Laboratory of High Field Laser Physics, Shanghai Institute of Optics and Fine Mechanics, Chinese Academy of Sciences, Shanghai 201800, China*
[2] *State Key Laboratory on Integrated Optoelectronics, College of Electronic Science and Engineering, Jilin University, Changchun 130012, China*
[3] *University of Chinese Academy of Sciences, Beijing 100049, China*
[4] *These authors contributed equally to this work.*

[†] *huailiang@jlu.edu.cn*

[*] *ya.cheng@siom.ac.cn*

[#] *zzxu@mail.shcnc.ac.cn*

**Abstract**

Recently, Yao et al. demonstrated the creation of coherent emissions in nitrogen gas with two-color (800 nm + 400 nm) ultrafast laser pulses [New J. Phys. **15**, 023046 (2013)]. Based on this two-color scheme, here we report on systematic investigation of temporal characteristics of the coherent emission at 391 nm ($N_2^+$: $B^2\Sigma_u^+(v=0)$ - $X^2\Sigma_g^+(v=0)$) by experimentally examining its evolution with the increase of the plasma channel induced by the intense 800 nm femtosecond laser pulses at a nitrogen gas pressure of ~25 mbar. We reveal unexpected temporal profiles of the coherent emissions, which show significant superradiance signatures owing to the quantum coherence via cooperation of an ensemble of excited $N_2^+$ molecules. Our findings shed more light on the mechanisms behind the laser-like emissions induced by strong-field ionization of molecules.

# Ⅰ. INTRODUCTION

By launching intense femtosecond laser pulses into a molecular gas, many interesting phenomena, such as high-order harmonic generation [1,2], above-threshold ionization and dissociation [3,4], bond softening and hardening [5,6], rotational excitation and molecular alignment [7], filamentation [8], supercontinuum generation [9], may occur. Thanks to these spectacular findings, strong field molecular physics has become one of the frontiers of contemporary physics and triggered many potential applications including molecular orbital imaging [10], coherent X-ray sources [11], attosecond chemistry [12,13], remote sensing [14,15], and so forth. Still, after nearly two decades of intensive investigations, new phenomena and new effects appear in this field. Recently, it has been observed that strong field ionization of molecules ($N_2$, $CO_2$, $H_2O$ etc.) can produce molecular ions mostly populated on the excited states, i.e. population inverted molecular ion systems are formed, leading to the formation of coherent emissions in molecular gases or even in remote air [16-24]. The underlying mechanism of the coherent emissions has not been completely identified, although our previous investigations have ruled out the possibilities of four-wave mixing and stimulated Raman scattering [25]. It has been known for a long time that narrow-bandwidth coherent emissions from cavity-less population inverted systems could have several closely-related but substantially different origins, including Dicke superradiance, superfluorescence, amplified spontaneous emission (ASE) and seed amplification (SA). The major difference in these processes is whether there is coherence in the atoms or molecules. For superradiance and superfluorescence to occur, such coherence is needed; whereas for ASE and SA such requirement is unnecessary [26-29]. To discriminate among the possible mechanisms behind the coherent emissions from the strong field ionized molecular gases, one approach is to examine its temporal evolution as a function of the length of gain medium, e. g., the plasma channel induced in the molecular gas by femtosecond laser pulses.

In this work, we simultaneously measure, by use of a two-color pump-probe method as described in [30], the temporal profiles of the 400 nm seed pulses and those of the coherent emissions at 391 nm wavelength (corresponding to a transition between $N_2^+$: $B^2\Sigma_u^+(\nu=0)$ - $X^2\Sigma_g^+(\nu=0)$) during their propagation in the nitrogen plasma channel produced by strong field ionization at a pressure of ~25 mbar. Surprisingly, we observe unexpected temporal oscillations in the 391 nm coherent emission pulses produced with the plasma channels of sufficient lengths. The information provides a strong evidence to exclude the seed amplification mechanism. On the other hand, such temporal oscillations can be reasonably regarded as signatures of the coherence in the molecular ions in the plasma channel. Thus, the origin of the narrow-bandwidth coherent emissions observed in our present experiment should be attributed to superradiance triggered by the 400 nm probe pulse.

## Ⅱ. EXPERIMENTAL SETUP

The experimental setup is sketched in Fig. 1, which is similar to that used in [30], but with a modification where the plasma channel length can be controlled by two movable thin steel plates mounted in a gas chamber. With these two movable plates, the length of the gas cell (and consequently the plasma channel length) can be continuously changed. Briefly, femtosecond laser pulses (1 kHz, 800 nm, 40 fs) from a commercial Ti:sapphire laser system (Legend Elite-Duo, Coherent, Inc.) were split into three arms. The first one with energy of 2.2 mJ was used as a pump, by using a 40 cm focal lens, to produce a plasma channel in the gas chamber filled with nitrogen gas at a pressure of ~25 mbar. The background pressure of the chamber was ~$10^{-2}$ mbar. The second one after frequency doubled by a 100 μm BBO acted as a seed to stimulate the 391 nm coherent emission. The polarization of seed pulse was set to be parallel to the pump pulse. The third arm was used to perform a cross-correlation measurement by generating the sum frequency with the seed pulse or the coherent 391 nm emission generated from the gas chamber in a 2-mm BBO crystal. The delay of the pulses from the three arms can be adjusted by the delay stages placed in the three laser beam paths. The sum frequency signal was recorded by a spectrometer (Shamrock 303i, Andor) with a 1200 grooves/mm grating. A Glan-Taylor prism was put before the spectrometer to measure the polarization property of 391 nm emissions.

## III. RESULTS AND DISCUSSION

Figure 2 shows a typical spectrum (black solid line) of strong narrow-bandwidth emissions at 391 nm produced from the $B^2\Sigma_u^+(\nu=0)$ - $X^2\Sigma_g^+(\nu=0)$ transition of $N_2^+$ in the plasma channel by simultaneously sending the 800 nm pump and the 400 nm seed into the nitrogen gas chamber with the delay time of $\Delta t = 0$ between the two pulses. During the experiment, by examining the far field beam profile, no diffraction by the edges of the holes on the two steel plates drilled by the pump laser could be observed. In addition, it is noteworthy that the gas pressure chosen in our experiment is 25 mbar. With such a low gas pressure, self-focusing will hardly occur (i.e., no filamentation and formation of energy reservoir [31]). Thus, the large-sized hole (diameter: ~1 mm) will allow nearly ~100% of the laser beam to pass through. As a consequence, the holes will not limit the propagation of the beam at our experiment. The data were averaged over 500 laser shots. The laser energies of the seed and pump pulses were 50 nJ and 2.2 mJ, respectively. For comparison, the spectra of the 400 nm seed alone (blue dash line) and the 800 nm pump alone (in the same spectral range around 400 nm, red dot line) were also recorded in the forward propagation direction after the chamber, as shown in Fig. 2. Importantly, it can be seen that when the 400 nm seed was blocked, the narrow-bandwidth emission disappeared in the spectrum produced by the 800 nm alone. Even more, it can also be noted from Fig. 2 that the signal intensity of the 400 nm seed, which is of only ~50 nJ pulse energy and whose spectrum is mainly peaked around ~400 nm with a bandwidth of ~4 nm, is very weak in the spectral region of the $B^2\Sigma_u^+(v=0)$ - $X^2\Sigma_g^+(v=0)$ transition of $N_2^+$ around 391 nm as shown by the blue dashed curve. However, the weak seed pulse can still play an important role in the generation of the 391 nm emission (black solid line). This is difficult to understand from the seed amplification point of view as for a strong amplification to occur, the seed pulse must be intense enough to induce the stimulated amplification efficiently. In addition, the 391 nm emission is measured to have a linear polarization with the direction parallel to that of the 400 nm seed, as demonstrated by the inset of Fig. 2. This observation clearly excludes the ASE mechanism for the 391 nm generation since ASE should show an isotropic polarization property [18].

To identify the role of the 400 nm seed, we measured the pulse shapes of coherent 391 nm emissions together with the pulse shape of the seed pulses as a reference by the cross-correlation method. Figures 3(a)-3(f) show the pulse shapes (blue line with circles) of the 391 nm emissions measured respectively with the plasma channel lengths of $l = 0, 2, 3, 4, 5$ and 6 mm along the propagation direction, which were achieved by truncating the plasma channel using the two movable plates. Note that the length of $l = 0$ mm represents the starting position where the 391 nm signal can merely be observable. The simultaneously measured pulse shapes (red solid line) of the 400 nm seeds (~100 fs) were also illustrated in Figs. 3(a)-3(f) along with those of the 391 nm coherent emission. It can be seen in Fig. 3 that all the measured 'characteristic durations' (which mean the durations of strongest main pulses as indicated in Fig. 3 by the arrows and defined by Eq. (1) below) of the 391 nm coherent emissions are in the ps range, which are much longer than that of the fundamental 800 nm pulse (~40 fs) for the sum frequency generation in our cross-correlation measurements. Thus these pulse shapes presented in Fig. 3 indeed reflect the temporal characteristics of the 391 nm coherent emissions.

One feature observed is that the 'characteristic duration' of the 391 nm emission becomes shorter as the plasma length becomes longer, which is also plotted in Fig. 4(a). In addition, signal oscillations appear in the pulse shape when the plasma length increases as shown in Fig. 3(d)-3(f), which is totally unexpected and is difficult to be explained with the seed amplification mechanism. Our analysis below shows that these features are signatures of the superradiance origin of the 391 nm coherent emission due to the quantum coherence via cooperation among ensemble excited molecules, which is triggered by the seed pulse [29, 32]. In such a case since the superradiance is extremely sensitive to the external injection field, the strong effect of the weak seed pulse could be justified.

Following the theoretical treatment of superradiance generation from a population inverted system [33], the characteristic duration of the superradiant emission, $\tau_R$, is directly related to the length of coherent population-inverted medium as follow,

$$\tau_R = \frac{8\pi\tau_{sp}}{3\rho\lambda^2 l} \qquad (1)$$

where $\tau_{sp}$ is the spontaneous decay time of the excited state, $\lambda$ the superradiant emission wavelength, $\rho$ the number density of the excited molecules, and $l$ the length of the medium (the plasma channel length). It can be seen from Eq. (1) that $\tau_R$ is inversely proportional to $l$, which qualitatively agrees well with our observation. We plot the characteristic duration of the superradiant emission as a function of the plasma length in Fig. 4(a), which clearly shows that the characteristic duration decreases as the plasma channel length increases. As indicated by the solid line in Fig. 4(a), the experimental data can be well fitted with the relation $\tau_R = C/(l-l_0)$, where $C$ and $l_0$ are constants determined by our experiment.

Moreover, the unexpected observation of the significant signal oscillations in the 391 nm pulse [Figs. 3(d)-3(f)] can also be understood following the framework of the seed-triggered superradiance. As the plasma length, $l$, increases, the radiant pulse length, $L_p = c\tau_R$ ($c$ is the light velocity) becomes shorter as we have discussed above. When $l > L_p$, the propagation effect is present for the collective emissions and the coherent molecular ion system will undergo the re-absorption and re-emission processes, leading to the signal oscillations in the temporal domain [29]. In our case, as shown in Fig. 4(a), when $l > 3.5$ mm, the characteristic durations keep the almost constant value of $\tau_R \sim 1$ ps, corresponding to $L_p = 0.3$ mm, satisfying the condition of $l > L_p$, and thus leading to the significant oscillations. However, for the longer pulse duration of 391 nm emission generated at the beginning of the plasma channel, the characteristic duration is $\tau_R \sim 8.3$ ps, corresponding to a pulse length of $L_p = 2.5$ mm in its propagation direction, which is longer than $l$ (< 1 mm), the oscillations are thus strongly reduced, as shown in Fig. 3(a). It should also be noted that when the plasma length becomes longer, a narrow peak appears at ~8.2 ps in Figs. 3(d)-3(f). This should be attributed to the revival of nitrogen molecular ions (the revival period of molecular ion is ~8.0 ps), which can also lead to the temporal modification of the signal intensity. It is also noteworthy that in our observation in Fig. 3(d)-3(f), the oscillation period of the 391 nm emission decreases with the increasing plasma channel length, which is consistent with the feature of superradiance as described in [29].

It is also well known that in the ideal case, the superradiance intensity should be proportional to $N^2$ [33], where $N$ represents the number of excited coherent atoms or molecules. Therefore, we measured the intensity of the 391 nm emission as a function of the gas pressure $P$ as shown in Fig. 4(b) (blue circle). In this measurement, the plasma length is kept to be 0.5 mm to avoid propagation effects such as reabsorption. It can be seen in Fig. 4(b) that the signal intensity can be well fitted with $I \sim C \cdot (P-P_0)^2$ (solid red line). Since the gas pressure $P$ is proportional to the number of excited molecules $N$, the measurement shown in Fig. 4(b) clearly follows the $N^2$ criteria, which provides an additional evidence for the superradiance.

## Ⅳ. CONCLUSION

To summarize, we have systematically investigated the temporal profiles of the 391 nm coherent emission from the $B^2\Sigma_u^+(\nu=0)$ - $X^2\Sigma_g^+(\nu=0)$ of $N_2^+$ as the evolution of the nitrogen plasma channel induced by strong field ionization. An unexpected signal oscillation was observed in the pulse shape of the 391 nm coherent emission, which suggests that the coherent emission may originate from the triggered superradiance due to the collective emission of coherent molecular ions rather than seed-injected amplification. This is also supported by the typical feature of the characteristic duration shortening with the increase of the interaction length. The new finding provides a possibility to the coherent manipulation of molecules in strong laser field.

## ACKNOWLEDGEMENTS

This work is financially supported by National Basic Research Program of China (Grant 2011CB808102, 2014CB921300), National Natural Science Foundation of China (Grant Nos. 11127901, 11134010, 61221064, 61235003, 11074098, 61275205 and 11204332), the Open Fund of the State Key Laboratory of High Field Laser Physics (SIOM), the Program of Shanghai Subject Chief Scientist (11XD1405500) and the Fundamental Research Funds of Jilin University.

**Captions of figures:**

Fig. 1 (Color online) Schematic diagram of the experimental setup.

Fig. 2 (Color online) Forward spectra measured with both the 800 nm pump and the 400 nm seed pulses (solid black line), the 800 nm pump pulse alone (dotted red line) and the 400 nm seed pulse alone (dashed blue pulse), respectively. Inset: the polarization property of the 391 nm emission measured with a Glan-Taylor prism (blue circles) and the fit with a square of sinusoidal function (red line).

Fig. 3 (Color online) Time-resolved cross-correlation signals of the 800 nm with the 400 nm seed pulse (solid red line) and the 391 nm coherent emission (blue line with circles) for the plasma channel lengths of (a) 0 mm, (b) 2mm, (c) 3 mm, (d) 4 mm, (e) 5 mm, (f) 6mm, respectively.

Fig. 4 (Color online) (a) The measured characteristic duration, $\tau_R$, of the 391 coherent emission as a function of the plasma channel length, $l$, (circles) and the fit with $\tau_R \sim C/(l-l_0)$ (solid line). (b) The measured 391 nm intensity as a function of the gas pressure (blue circles) and the fit with $I \sim C \cdot (P-P_0)^2$ (red-solid line).

**Figure 1:**

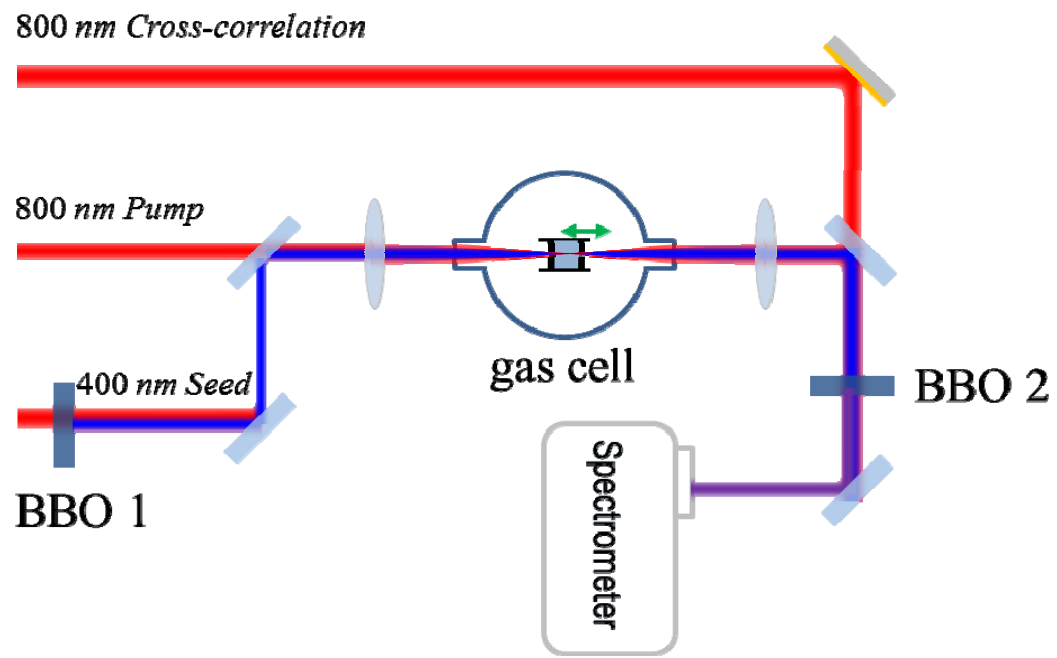

**Figure 2:**

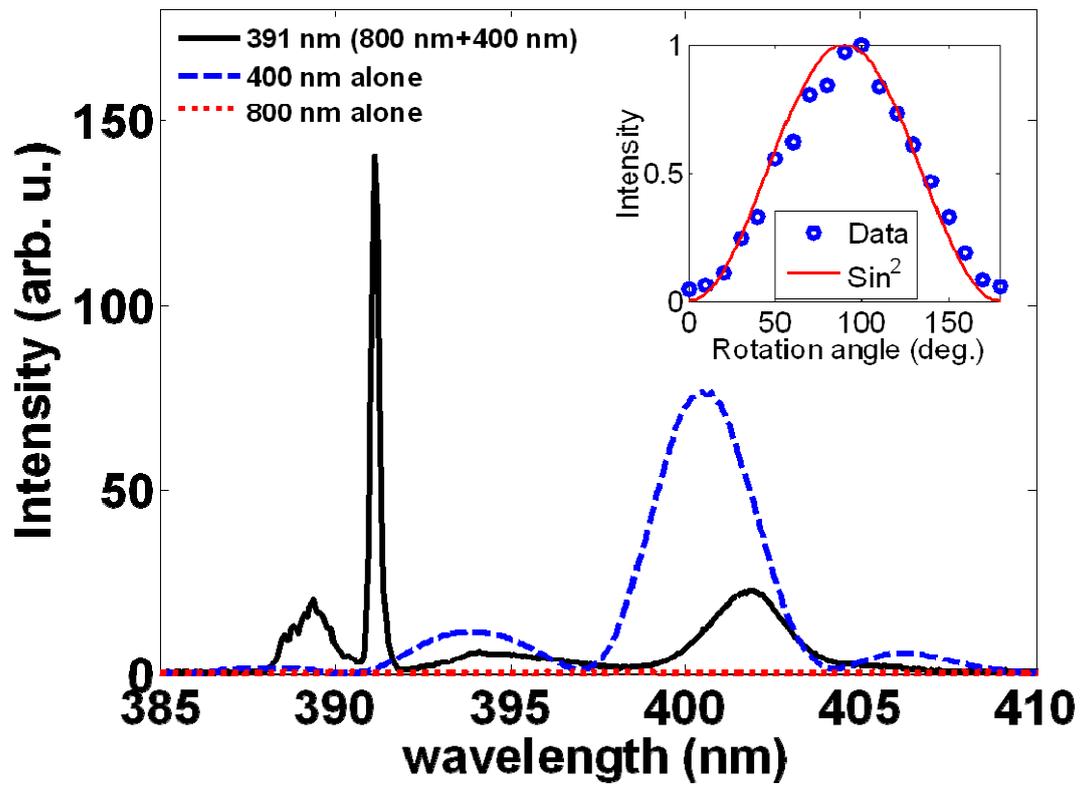

**Figure 3:**

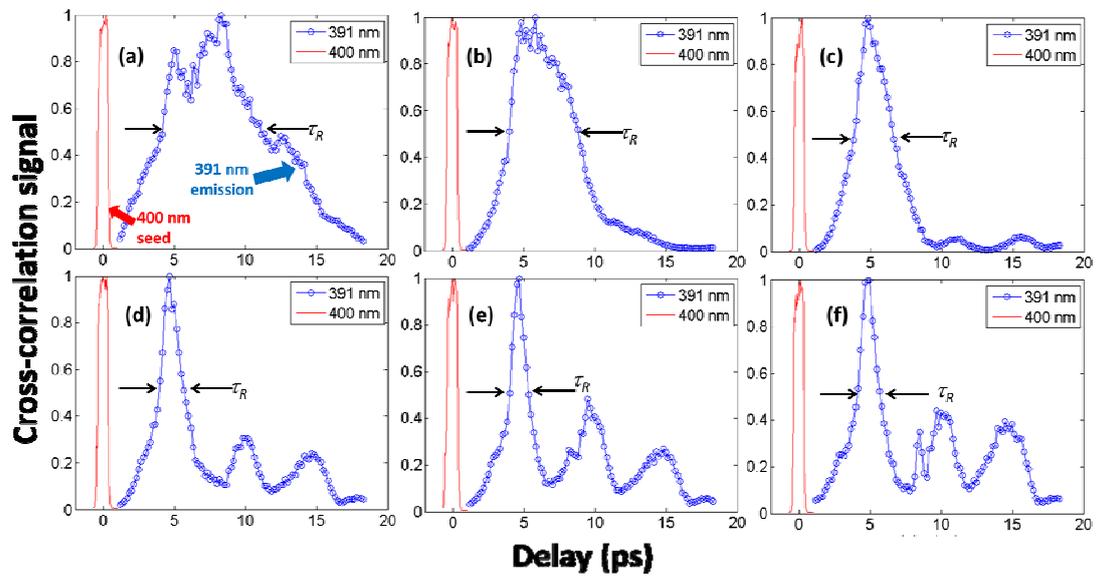

**Figure 4:**

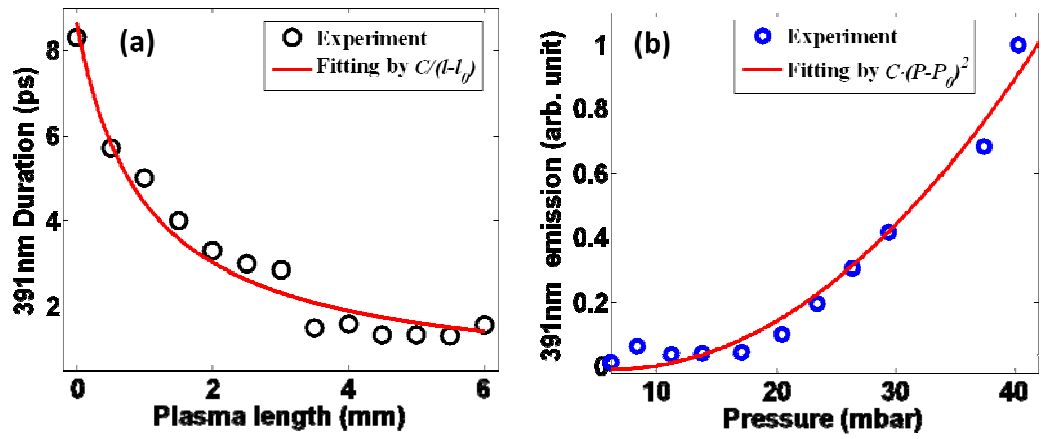